\documentclass[12pt]{article}
\pdfoutput=1

\usepackage{amsmath,amssymb,amsthm,amstext,mathrsfs,bbm}
\usepackage{bbold}
\usepackage{latexsym,amscd,amsbsy,dsfont,amsfonts}
\usepackage{graphicx}
\usepackage[pdftex,colorlinks=true]{hyperref}
\usepackage{array}
\usepackage{wrapfig}

\usepackage{placeins}
\newcommand{\flush}{\FloatBarrier}

\newcolumntype{L}{>{$}l<{$}}
\newcolumntype{C}{>{$}c<{$}}
\newcolumntype{R}{>{$}r<{$}}

\newcommand{\be}{\begin{equation}}
\newcommand{\ee}{\end{equation}}
\newcommand{\bea}{\begin{eqnarray}}
\newcommand{\eea}{\end{eqnarray}}
\numberwithin{equation}{section}

\begin{document}

\baselineskip 17pt

\begin{titlepage}

\vfill

\begin{flushright}
\end{flushright}

\vfill
\begin{center}
	{\Large\bf  Short-lived modes from hydrodynamic dispersion relations}
  	\vskip 1.5cm
      	Benjamin Withers\\
   	\vskip .6cm
	benjamin.withers@unige.ch\\
	\vskip .6cm
        \begin{small}
      	\textit{Department of Theoretical Physics, University of Geneva, 24 quai Ernest-Ansermet, 1214 Gen\`eve 4, Switzerland}
        \end{small}\\*[.6cm]
\end{center}
\vfill

\begin{center}
\textbf{Abstract}
\end{center}
We consider the dispersion relation of the shear-diffusion mode in relativistic hydrodynamics, which we generate to high order as a series in spatial momentum $q$ for a holographic model. We demonstrate that the hydrodynamic series can be summed in a way that extends through branch cuts present in the complex $q$ plane, resulting in the accurate description of multiple sheets.
Each additional sheet corresponds to the dispersion relation of a different non-hydrodynamic mode.
As an example we extract the frequencies of a pair of oscillatory non-hydrodynamic  black hole quasinormal modes from the hydrodynamic series.
The analytic structure of this model points to the possibility that the complete spectrum of gravitational quasinormal modes may be accessible from the hydrodynamic derivative expansion.

\vfill

\end{titlepage}
\setcounter{equation}{0}

\section{Introduction}
As an effective theory of conserved currents near equilibrium, the strength of hydrodynamics lies in its universal applicability. This is reflected in the fact that at any given order in the derivative expansion only finitely many terms can be written down, reducing any theory with the appropriate limit to a set of transport coefficients.

Recently the convergence properties of the hydrodynamic derivative expansion have been investigated. In the context of boost-invariant flows in various models, it has been shown that the late proper time expansion of the stress tensor has a zero radius of convergence \cite{Heller:2013fn}, with many further interesting investigations in this context \cite{Heller:2015dha,Basar:2015ava,Aniceto:2015mto,Florkowski:2016zsi, Denicol:2016bjh, Heller:2016rtz, Florkowski:2017olj, Spalinski:2017mel, Casalderrey-Solana:2017zyh, Heller:2018qvh}. The divergent nature of the series was argued to be connected to the factorial growth in the number of allowed terms at each order in the hydrodynamic expansion. These works demonstrated that the perturbative expansion can be understood as one piece of a more general transseries, containing non-perturbative contributions from short-lived non-hydrodynamic modes which decay exponentially in proper time. The contribution of these non-hydrodynamic modes were inferred from the perturbation series using the machinery of Borel-Pad\'e. Other examples are provided by the summation of the hydrodynamic expansion applied to a cosmological setting \cite{Buchel:2016cbj}, and to the retarded correlation functions in kinetic theory \cite{Kurkela:2017xis}. See also the recent review for related results \cite{Aniceto:2018bis}.

It is an interesting question in general, as to how much information about the underlying microscopic theory can be extracted from a hydrodynamic derivative expansion. In this work we adopt a different approach to this question, focussing on the large order behaviour of the dispersion relations of hydrodynamic modes. This allows us to directly access short-lived non-hydrodynamic mode dispersion relations of the underlying theory, by appropriately summing the hydrodynamic series. Aside from direct access to dispersion relations, an advantage to this approach is that we are able to elucidate connections that exist between various modes through navigating the complex $q$ plane. On a practical level, working with linear hydrodynamics is a crucial simplification which allows us to reach high orders in the hydrodynamic expansion, in contrast with previous approaches where a simplification came from a choice of symmetries of physical scenario. 

By way of a concrete example, we focus our attention on the dispersion relation of the shear-diffusion mode, whose hydrodynamic expansion can be written in the form of a series in spatial momentum $q$,
\be
\omega(q) = -i D q^2 + O(q)^4. \label{sheardiff}
\ee
Note that $\omega(0) = 0$ as required for a hydrodynamic mode; at $q=0$ it is a zero mode associated to translational symmetry and momentum conservation. The leading term in $q^2$ can be obtained from first-order relativistic hydrodynamics, as the dispersion relation that arises for transverse linear perturbations around equilibrium, finding,
\be
D = \frac{\eta}{\varepsilon + p},
\ee
where $\eta$ is the shear viscosity, $\varepsilon$ the energy density and $p$ the pressure.\footnote{A review of this material and relativistic hydrodynamics in general can be found in \cite{Kovtun:2012rj}.} We may also include a finite charge density, $n$, with chemical potential $\mu$ and the hydrodynamic variables become $T,\mu$ and a fluid velocity.

The coefficients of this series  \eqref{sheardiff}  are calculable once an equation of state $p(T,\mu)$ and a set of transport coefficients (at first order this is just $\eta$) are given. This data may be fixed by matching to an underlying microscopic model if it has a hydrodynamic limit. In this paper we pick an underlying model and generate the expansion \eqref{sheardiff} in precisely this way. A particularly convenient theory to work with is one with a holographic dual -- then it is relatively straightforward, at least numerically, to obtain higher order terms in the expansion \eqref{sheardiff} for hydrodynamics in $d$ dimensions by solving ODEs in a spacetime of dimension $d+1$. In holography the first term has been computed exactly and the associated transport coefficient found to be, famously, \cite{Policastro:2001yc,  Kovtun:2004de}
\be
\eta = \frac{s}{4\pi}
\ee
where $s$ is the entropy density. The Taylor expansion \eqref{sheardiff} proceeds in even powers of $q$ and one must go to third order in the hydrodynamic expansion to extract the next correction. For holography at zero charge density in $d=4$ this was carried out in \cite{Baier:2007ix}, for the sound mode at third order see \cite{Grozdanov:2015kqa}.

In this paper we pick the specific holographic model defined by Einstein-Maxwell theory in AdS$_{4}$. In this case the equilibrium state is given by a Reissner-Nordstrom black brane in the bulk. The linear modes of this theory, as poles of retarded correlators of the conserved currents, are quasinormal modes of this black brane. These modes exhibit a rich and interesting structure as a function of $q/\mu$ and $T/\mu$, as was depicted beautifully in  \cite{Brattan:2010pq}. Since we are first interested in the hydrodynamic Taylor series and its convergence properties, we are interested in the analytic structure in the complex $q$ plane. The exact\footnote{i.e. without performing the hydrodynamic expansion} $\omega(q)$ has a multi-sheeted structure, and we shall see that the longest-lived sheet (that which has the largest value of Im$\omega(0)$, the hydrodynamic sheet) contains branch points at 
\be
q = \pm i q_\ast,\qquad \text{where}\qquad q_\ast \equiv \frac{1}{2\mu}\sqrt{\frac{\varepsilon+p}{D}} = \frac{\varepsilon+p}{2\mu\sqrt{\eta}}. \label{qast}
\ee
At least for the examples of $T/\mu$ that we study, these branch points are the closest non-analyticity to $q=0$, and set a finite radius of convergence of the hydrodynamic expansion of the dispersion relation, \eqref{sheardiff}. This point, and thus the physical origin of the scale $q_\ast$, can be interpreted as the location of a collision between a hydrodynamic and non-hydrodynamic mode on the imaginary $q$ axis. Circumnavigating this point, we can move to a second, shorter-lived sheet, and that sheet is associated to a non-hydrodynamic mode. 

The key result of this work is that we can first determine the location of these branch points from hydrodynamic series itself, and then we can accurately describe such additional sheets with a particular summation of the hydrodynamic expansion. In doing so, we tease out the dispersion relation of non-hydrodynamic modes from hydrodynamic data.\footnote{In the context of divergent series the non-hydrodynamic mode arises as from an ambiguity in performing the inverse Borel transformation, see e.g. \cite{Heller:2013fn, Heller:2015dha}.} 

The remainder of the paper is structured as follows. In section \ref{model} we detail the underlying model we will use to generate the hydrodynamic expansion of the shear-diffusion mode to high order in $q$. In section \ref{convergence} we explore the convergence of the hydrodynamic expansion, and in \ref{extending} we extend the expansion onto a second sheet by circumnavigating the branch points, extracting the frequency of a pair of non-hydrodynamic modes. We conclude with a discussion in section \ref{discussion}.

\section{Hydrodynamic coefficients from holography \label{model}}
The spectrum of vector quasinormal modes of a Reissner-Nordstrom black brane in Einstein-Maxwell theory in AdS$_4$, contains a dispersion relation of the form \eqref{sheardiff} at small $q$. This is because the model is holographically dual to a CFT at finite $T$ and $\mu$, and this mode is precisely the shear diffusion mode in the hydrodynamic limit of that CFT. The goal of this section is to compute the small-$q$ Taylor expansion of the dispersion relation of this mode to high order, by solving an appropriate linear gravitational problem on the Reissner-Nordstrom background.

In Schwarzschild-like coordinates we have the following Reissner-Nordstrom black brane metric and gauge field, with holographic coordinate $z$ scaled such that the horizon is at $z=1$,
\bea
g &=& \frac{1}{z^2}\left(-f(z) dt^2 + \frac{dz^2}{f(z)} + dx^2 + dy^2 \right)\\
A &=& \mu\left(1- z\right) dt
\eea
where the metric function
\be
f(z) = 1 - \left(1+\frac{\mu^2}{4}\right) z^3 + \frac{\mu^2 z^4}{4}.
\ee
On this background we consider the following transverse metric and gauge field linear perturbations, 
\bea
\delta g(t,z,x,y) &=& \frac{e^{-i \omega t + i q x}}{z^2}\left(h_{ty}(z)2dt dy + h_{xy}(z)2dx dy\right),\\
\delta A(t,z,x,y)  &=& e^{-i \omega t + i q x} h_y(z) dy,
\eea
which satisfy a closed set of ODEs in the $z$-coordinate of total differential order $5$. It is possible to construct two linear combinations of the above variables, the master fields $\phi_\pm(z)$, which satisfy decoupled ODEs \cite{Kodama:2003kk}. Here we define,
\bea
\phi_\pm(z) = \pm i \frac{-q f h_{xy}' + c_\mp z \omega h_y}{(c_--c_+) z \omega q^2}
\eea
where
\be
c_\pm = \frac{3(4+\mu^2) \pm \sqrt{9 (4+\mu^2)^2 + 64  \mu^2q^2}}{8\mu}.
\ee
With this change of variables we have the following equations of motion for the master fields, 
\be
-f \left(f \phi_\pm'\right)' + f(q^2 + \mu^2 z^2 - \mu z c_\pm) \phi_\pm - \omega^2 \phi_\pm = 0. \label{master}
\ee
The remaining equation is an equation that determines $h_{ty}$, but it is dependent on $\phi_{\pm}$ and so for the purposes of finding the values of $\omega$ we do not need to consider it. 

Note that this master field equation reveals something about the analytic structure of solutions we will obtain, since it contains a square root involving $q$. The two sectors $\phi_\pm$ are therefore connected through branch cuts, with a corresponding branch point occurring at \eqref{qast}.
However, the purpose of the holographic model for this work is purely to compute the hydrodynamic dispersion relation \eqref{sheardiff} as a Taylor series in $q$, and so we are strictly only interested in the hydrodynamic sheet in this section. This is the sheet whose frequency satisfies $\omega(0) = 0$, and corresponds to the `$+$' choice above.

We impose boundary conditions for quasinormal modes, namely the absence of sources on the boundary of AdS at $z=0$, and ingoing boundary conditions at the black hole horizon at $z=1$. We therefore have a boundary value problem to solve, with eigenvalues $\omega^2$ which will depend on the momentum of interest $q$.  To compute the hydrodynamic $\omega$ as a Taylor series in $q$, we expand both the perturbation $\phi_+$ and the eigenvalue $\omega$, resulting in an infinite series of ODE boundary value problems, i.e.
\bea
\phi_{+}(z) &=& \sum_{n=0}^{\infty} \psi_n(z) q^{2n},\\
\omega(q) &=& \sum_{n=1}^{\infty} \omega_n q^{2n}.
\eea
Notice that the expansion for $\omega(q)$ begins at $q^2$ in accordance with \eqref{sheardiff}, selecting the hydrodynamic mode.
The ingoing behaviour of $\phi_{+}$ near the horizon can be expressed as a series around $z=1$,
\be
\phi_+(z) = (1-z)^{-\frac{i \omega}{4\pi T}}\left(1 
+ \left(
\frac{12i(\mu^2-4)\omega}{(\mu^2-12)^2} - \frac{4(q^2 + \mu(\mu-c_+))}{\mu^2 - 12 + 8 i \omega}
\right)(1-z) 
+ O(1-z)^2\right)\nonumber
\ee
where we have fixed an arbitrary coefficient to unity. We subsequently expand this expression in $q$ to generate horizon boundary conditions for each of the $\psi_n$. At level $0$ we can easily solve the equation, to obtain $\psi_0(z) = z$. Next order will determine $\psi_1(z)$ and $\omega_1$, and so on. We turn to numerics to evaluate the remaining $\psi_n$ and associated frequency coefficients $\omega_n$ for $n\geq 1$.
To obtain $\omega_n$ one must solve $n$ ODEs. We do this using a shooting method, integrating out from the horizon with a guess for $\omega_n$ and iteratively improving this guess until we satisfy the boundary conditions at $z=0$. The values we obtain are summarised in table \ref{wtable}.

\section{Radius of convergence\label{convergence}}
\begin{figure}[h!]
\begin{center}
\includegraphics[width=0.65\columnwidth,clip]{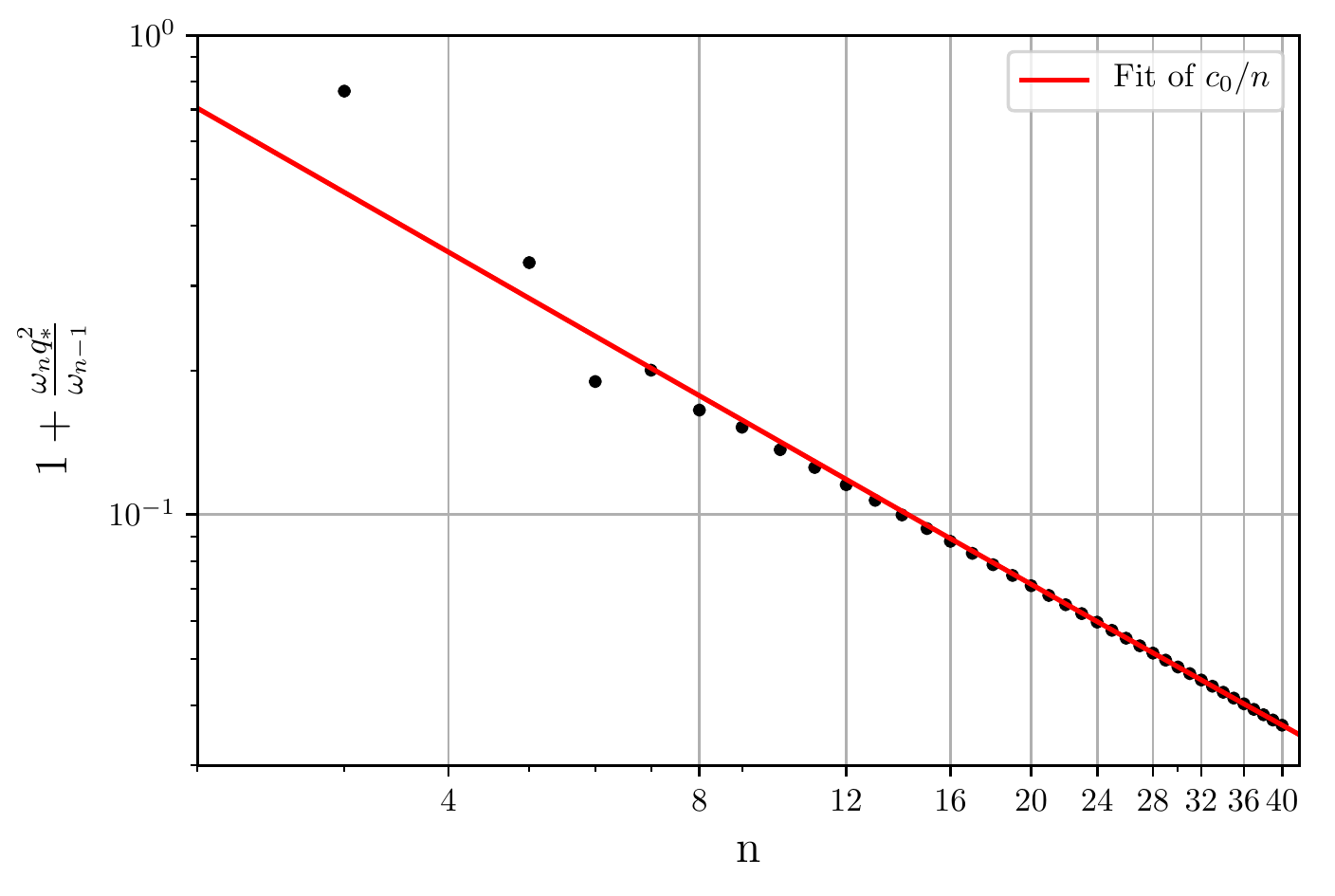}
\caption{The sequence $1+r_n$, for $r_n$ defined in \eqref{rn} as ratios of successive terms in the hydrodynamic expansion. On this log-log plot $1+r_n$ is shown to converge linearly to $0$, indicated by the red line whose slope is $-1$. This confirms that the convergence of the hydrodynamic expansion is set by the scale $q_\ast$ defined in \eqref{qast}. \label{converge}}
\end{center}
\end{figure}

Using the specific microscopic model detailed in the last section, we have determined the Taylor expansion of $\omega(q)$ (up to some finite order $N=40$, i.e. hydrodynamic order $79$),
\be
\omega(q) = \sum_{n=1}^{N}\omega_n q^{2n}. \label{taylorq}
\ee
Rendering the $\omega_n$ dimensionless using the scale $q_\ast$, we present the coefficients obtained in table \ref{wtable}.
\begin{table}
\begin{center}
\begin{tabular}{R|R|R|R|R|R|R|R}
n & -i \omega_n q_\ast^{2n-1} & n & -i \omega_n q_\ast^{2n-1} & n & -i \omega_n q_\ast^{2n-1} & n & -i \omega_n q_\ast^{2n-1} \\
\hline
1&-0.250&11&3.89\times 10^{-3}&21&1.55\times 10^{-3}&31&8.80\times 10^{-4}\\2&-7.07\times 10^{-2}&12&-3.44\times 10^{-3}&22&-1.45\times 10^{-3}&32&-8.40\times 10^{-4}\\3&1.67\times 10^{-2}&13&3.07\times 10^{-3}&23&1.36\times 10^{-3}&33&8.03\times 10^{-4}\\4&-1.69\times 10^{-2}&14&-2.76\times 10^{-3}&24&-1.28\times 10^{-3}&34&-7.69\times 10^{-4}\\5&1.12\times 10^{-2}&15&2.50\times 10^{-3}&25&1.20\times 10^{-3}&35&7.37\times 10^{-4}\\6&-9.09\times 10^{-3}&16&-2.28\times 10^{-3}&26&-1.14\times 10^{-3}&36&-7.07\times 10^{-4}\\7&7.27\times 10^{-3}&17&2.09\times 10^{-3}&27&1.08\times 10^{-3}&37&6.80\times 10^{-4}\\8&-6.07\times 10^{-3}&18&-1.93\times 10^{-3}&28&-1.02\times 10^{-3}&38&-6.54\times 10^{-4}\\9&5.15\times 10^{-3}&19&1.79\times 10^{-3}&29&9.69\times 10^{-4}&39&6.29\times 10^{-4}\\10&-4.44\times 10^{-3}&20&-1.66\times 10^{-3}&30&-9.23\times 10^{-4}&40&-6.06\times 10^{-4}  
\end{tabular}
\caption{
Coefficients of the shear-diffusion mode dispersion relation in a hydrodynamic gradient expansion \eqref{sheardiff}, at $\frac{T}{\mu} = \frac{1}{4\pi}$ in the holographic dual of Einstein-Maxwell in AdS$_4$. The scale $q_\ast$ is defined in \eqref{qast}. For the purposes of presentation we have truncated to three significant digits. These numbers, or at least their higher precision counterparts, contain the dispersion relation of at least two non-hydrodynamic modes, and extracting them is the purpose of this paper.\label{wtable}}
\end{center}
\end{table}
Next we can determine the radius of convergence by looking at the ratios of successive terms. Specifically we form the new dimensionless sequence, 
\be
r_n \equiv \frac{\omega_{n}q_\ast^2}{\omega_{n-1}} \label{rn}.
\ee
The large-$n$ behaviour of this sequence is $1+r_n = c_0/n$ as illustrated in figure \ref{converge}, so $r_n$ converges to $-1$ confirming that as expected, the radius of convergence of the hydrodynamic expansion is governed by $q^2 = -q_\ast^2$, i.e. the branch points at $q = \pm i q_\ast$.

\begin{figure}[h!]
\begin{center}
\includegraphics[width=0.8\columnwidth,clip]{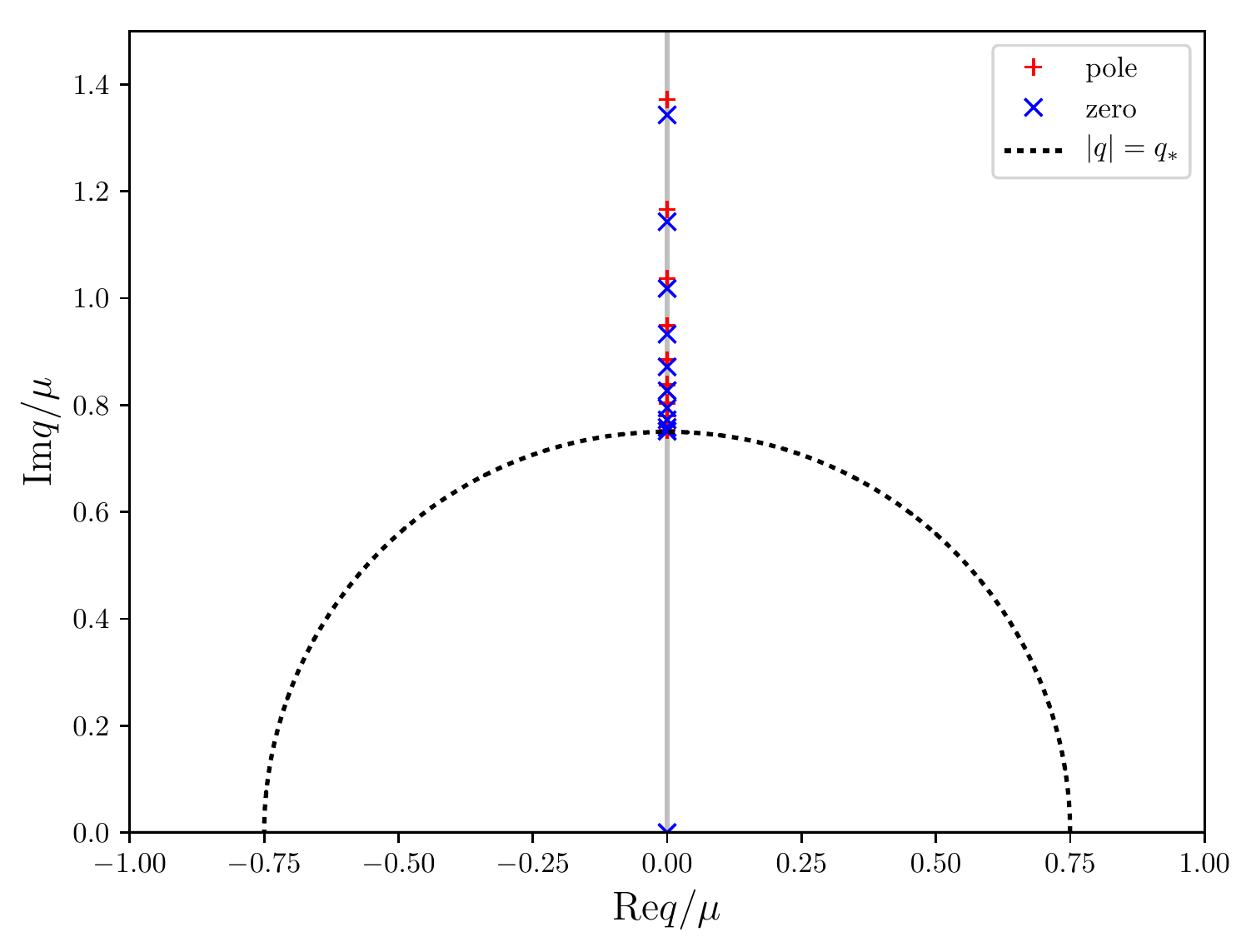}
\caption{Poles and zeros of the Pad\'e approximant, ${\cal P}_q$, to the hydrodynamic Taylor expansion of the shear-diffusion mode dispersion relation $\omega(q)$ in holography. The alternating sequence of poles and zeros along a radial ray is indicative of a branch point at $q = i q_\ast$, the closest non-analytic feature to the hydrodynamic limit $\omega = q=0$. \label{pade}}
\end{center}
\end{figure}

As a further diagnostic, we compute the diagonal Pad\'e approximant of the series. i.e. defining a ratio of two polynomials,
\be
{\cal P}_q(q) = \frac{\sum_{i=0}^{N} a_i q^i}{1+\sum_{j=1}^{N} b_j q^j} \label{P}
\ee
calculate the coefficients $a_i,b_i$ from the coefficients $\omega_n$ by matching order-by-order in the Taylor series around $q=0$. We plot the poles and the zeros of ${\cal P}_q(q)$ in figure \ref{pade} for the upper-half plane, showing an alternating sequence of poles and zeros along a radial ray. The closest pole to $q=0$ is given by $q \simeq 0.753i$, a good indicator of the branch point at $iq_\ast \simeq 0.750i$ (the same structure exists in the lower-half plane). 

\section{Through the branch cut\label{extending}}
As we have established in the last section the Taylor expansion \eqref{taylorq} has a finite radius of convergence set by a branch point in the complex-$q$ plane. We can determine the location of this point empirically from our hydrodynamic data in table \ref{wtable}, by looking at the radius of convergence, and also at the poles and zeros of the Pad\'e approximant.

\begin{figure}[h!]
\begin{center}
\includegraphics[width=1.0\columnwidth,clip]{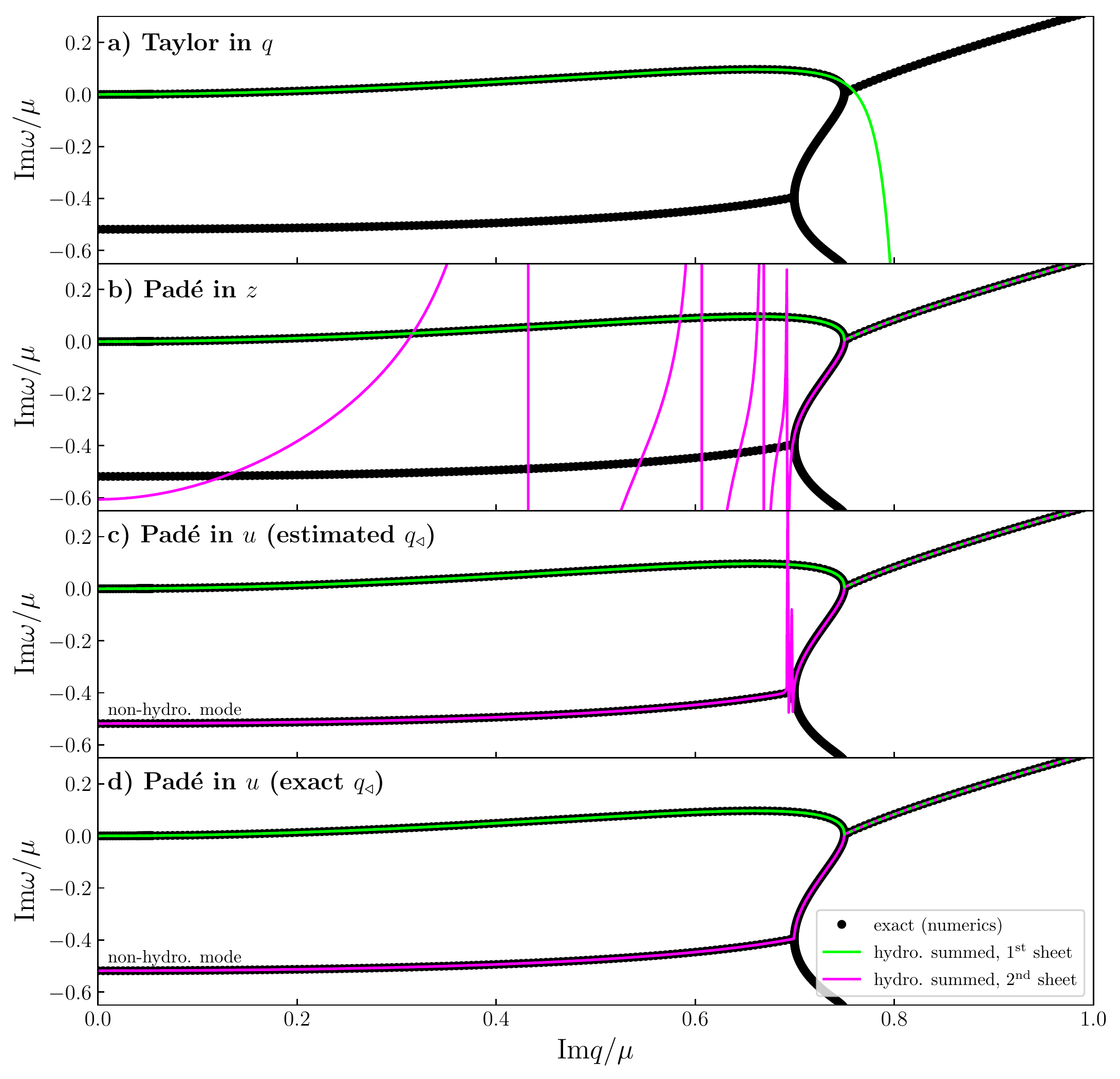}
\caption{Extending the hydrodynamic derivative expansion to a second sheet. The black points in all panels show the exact values, obtained numerically via a standard shooting calculation, and without invoking a derivative expansion. The coloured curves are as follows: \textbf{Panel a)} shows the hydrodynamic Taylor expansion, which converges only up to the branch point shown. \textbf{Panel b)} shows the hydrodynamic expansion summed as a Pad\'e approximant ${\cal P}_z$ of a complex variable $z$ as defined in \ref{zdef}, producing two sheets (shown in lime and magenta respectively) which show excellent agreement up to a second branch point. \textbf{Panels c) \& d)} show the Pad\'e approximant ${\cal P}_u$ of a complex variable $u$ as defined in \ref{udef}, extending past another branch point and allowing the summed hydrodynamic expansion to describe the real line. The error in the empirically inferred second branch point of panel c) is damped as the real line is approached. \label{quad}}
\end{center}
\end{figure}

Armed with this knowledge of the nearest non-analytic obstacle to convergence of the hydrodynamic expansion, we adopt a new variable, $z$, that can cover more than one sheet. In particular, the branch point is of square-root type and so we want to describe two sheets in the simplest possible way. Thus we are led to consider the following quadratic polynomial in z, 
\be
q^2 = a + b z + c z^2.
\ee
We fix the coefficients $a,b,c$ as follows. First we demand that $q^2=0$ at $z=0$ so we may consider instead a new Taylor series around $z=0$, which implies $a=0$. Next we require that the two solutions for $z$ have a branch point at a prescribed location, $q^2=(q^2)_{bp}$, which fixes $c = -\frac{b^2}{4(q^2)_{bp}}$. Finally we use the freedom to rescale $z$ to set $b=-2(q^2)_{bp}$, resulting in 
\be
\frac{q^2}{(q^2)_{bp}}= -z(2+z). \label{zdef}
\ee
For instance, $q=0$ on the first sheet is given by $z=0$ and on the second by $z=-2$.
Thus we are matching the topology of the surface, and utilising some geometrical information: the location of the branch point itself.\footnote{For conformal maps in the context of the Borel plane, see for example \cite{Jentschura}.} We set $q_{bp} = i q_\ast$, the branch point discussed in section \ref{convergence}. With the transformation in place, we convert the expansion \eqref{taylorq} to a new Taylor expansion in $z$ around $z=0$. In order to see that there are now two sheets, one must convert back to the variable $q$. However for this model we have found that we must go one step further, and sum the series as a Pad\'e approximant in order to achieve good results. Specifically we construct the diagonal Pad\'e approximant in the variable $z$, ${\cal P}_z(z)$. As can be seen in panel b) of figure \ref{quad} under comparison with the exact results, ${\cal P}_z(z)$ extends the hydrodynamic expansion onto a second sheet, with remarkable agreement.

\begin{figure}[h!]
\begin{center}
\includegraphics[width=0.8\columnwidth,clip]{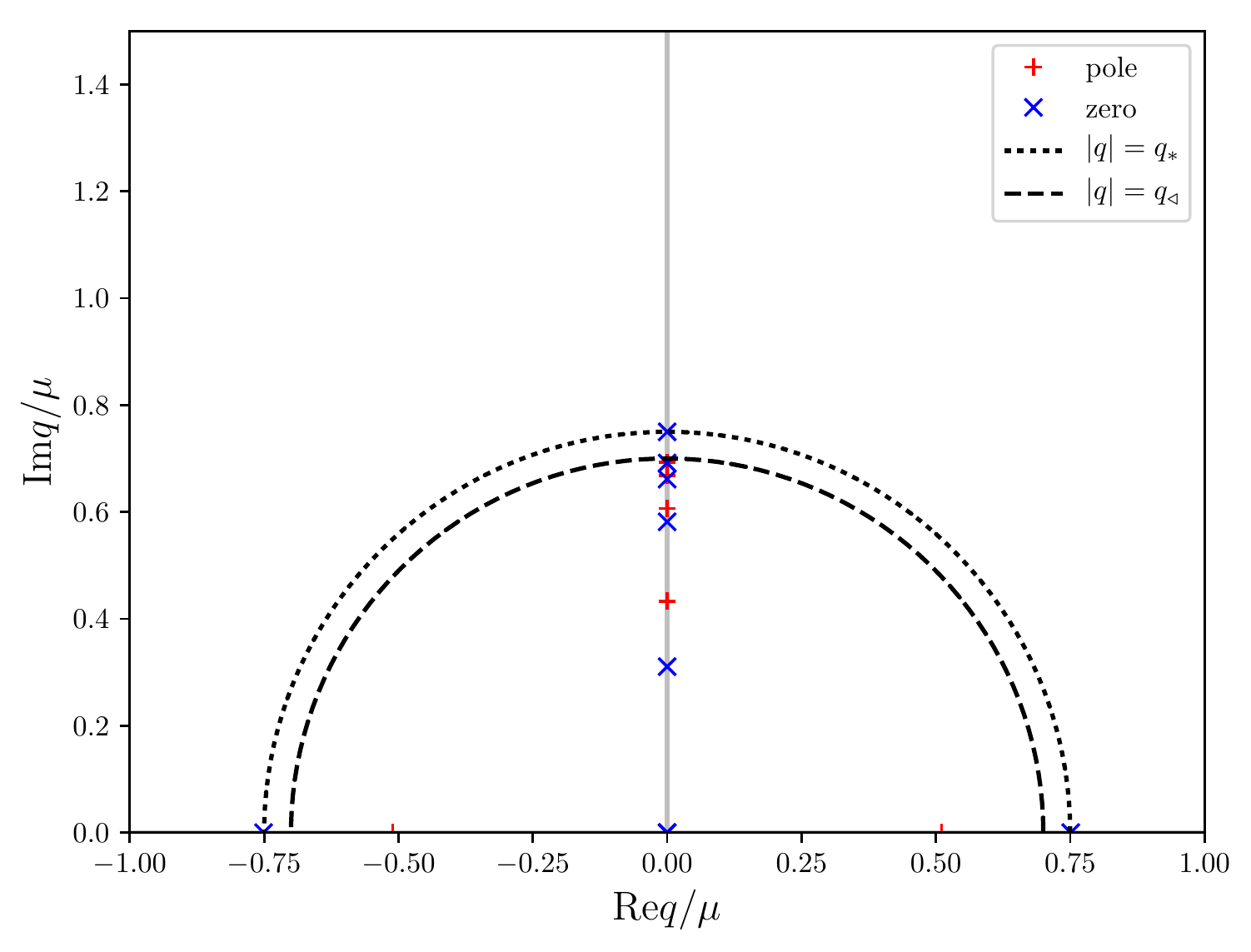}
\caption{Poles and zeros of the Pad\'e approximant, ${\cal P}_z$, which is constructed using the variable $z$ (defined in \eqref{zdef}) which covers more than one sheet. The outer dotted line corresponds to the original branch point discussed in section \ref{convergence}, whilst the inner dashed line corresponds to another branch point we must conquer to reach the real line, as indicated by the sequence of poles and zeros extending inwards from it, corresponding to the features in the magenta line, panel b) of figure \ref{quad}.\label{pade2}}
\end{center}
\end{figure}

Note however that there is a second branch point preventing us from accessing the real line on the second sheet, and ${\cal P}_z$ is not sufficient. This is also indicated by a sequence of poles and zeros of ${\cal P}_z$ in figure \ref{pade2}.
To pass this second branch point and get back to the real line, we simply repeat the above procedure transforming the variable $z$ to a new complex variable $u$,
\be
\frac{z}{z_{bp}} = - u(2+u), \label{udef}
\ee
so that we take into account the second branch point at $z=z_{bp}$. Note that we have now generated four sheets in total, though we will restrict our attention to the first three. As before, it is possible to estimate the value of $z_{bp}$ purely from the hydrodynamic data. Specifically, we look for the pole of ${\cal P}_z(z)$ which occurs after we pass onto the second sheet, in figure \ref{pade2}, and to the order considered we find the location of this pole $q = i q_\triangleleft \simeq 0.69i\mu$. To assess how accurate this hydrodynamic estimate is, we can compare it to the exact location of the branch point computed in holography, for which we find $iq_\triangleleft \simeq 0.70i \mu$.\footnote{In other words to obtain the branch point we solve the holographic QNM equations without invoking a hydrodynamic expansion and numerically obtain the value quoted.}
We show the Pad\'e approximant for the new variable $u$, ${\cal P}_u$ in panels c) and d) of figure \ref{quad}. In figure c) the variable $u$ is constructed using the estimated branch point location, $i q_\triangleleft \simeq 0.69i\mu$, whilst in panel d) the variable $u$ is constructed using the exact branch point location, $i q_\triangleleft \simeq 0.70i\mu$.
Whilst in panel c) we pay a penalty for not knowing the exact value of $q_\triangleleft$, remarkably it appears that the resulting deviations are restricted to the region near the branch point, decaying rapidly as we go towards $q=0$. We have verified that this behaviour also persists after varying $q_\triangleleft$ more significantly, and the value of the frequency on the real axis appears stable.

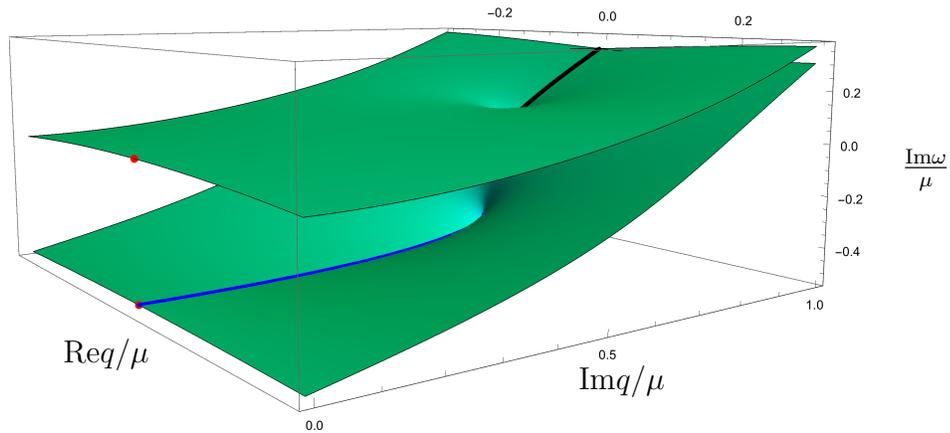
\begin{figure}[h!]
\begin{center}
\includegraphics[width=0.7\columnwidth,clip]{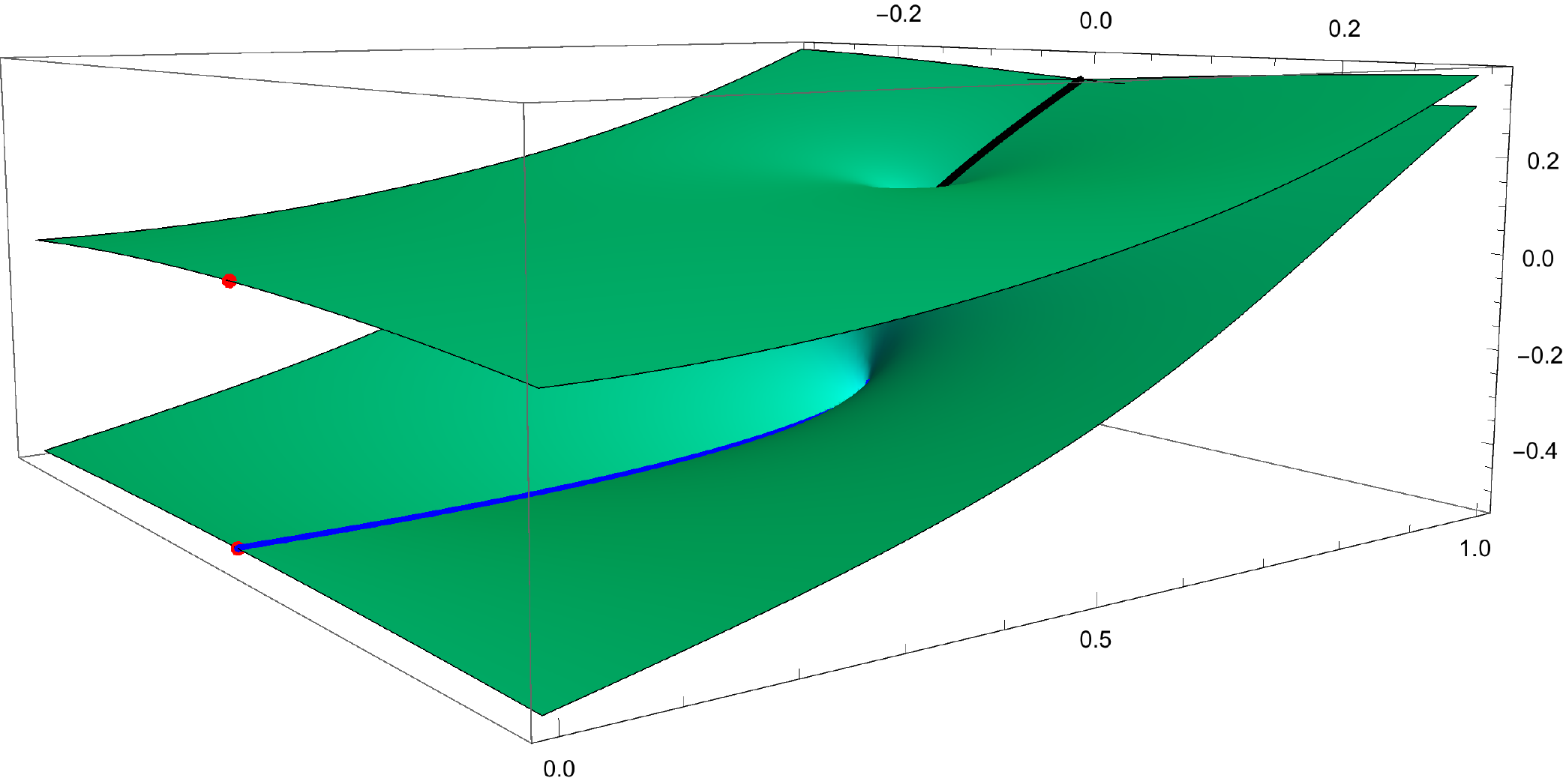}
\begin{picture}(1,1)(0,0)
\put(-95,20){\makebox(0,0){Im$q/\mu$}}
\put(-290,30){\makebox(0,0){Re$q/\mu$}}
\put(20,100){\makebox(0,0){$\frac{\text{Im}\omega}{\mu}$}}
\end{picture}
\caption{Im$\omega(q)$ for two sheets constructed from a summation of the hydrodynamic series for the dispersion relation of the shear-diffusion mode in holography. The upper red dot corresponds to the hydrodynamic point, $q=0$ where $\omega(0) = 0$ on that sheet. The lower red dot corresponds to a non-hydrodynamic mode at $q=0$, where $\omega(0)$ on that sheet corresponds to the frequency of a non-hydrodynamic mode, given in table \ref{nonhydrotable}. The solid blue and black lines correspond to branch cuts. A comparison with the exact result on the Im$q$ axis is given in figure \ref{quad}. \label{sheets}}
\end{center}
\end{figure}

A comparison is made in table \ref{nonhydrotable} of the first non-hydrodynamic mode frequency for various truncation orders, made using the estimated $q_\triangleleft$. In this table we see that by order 11 in the hydrodynamic expansion we have a reasonable estimate for the frequency, and by the time we reach order 79 the agreement is correct to four significant digits. Finally, in figure \ref{sheets} we present plots of the two sheets of $\omega(q)$ we have obtained in the complex $q$ plane from hydrodynamic data. This of course displays the dispersion relations of the non-hydrodynamic mode on the real-$q$ axis too, which we can effortlessly read off from the summed series.

\begin{table}[h!]
\begin{center}
\begin{tabular}{R|R|R}
\text{hydro. order} & \omega(0)/\mu \text{ (2\textsuperscript{nd} sheet)} & \omega(0)/\mu \text{ (3\textsuperscript{rd} sheet)}\\
\hline
3 & 0.0276 + 0.2475i & -0.0276 + 0.2475i\\
5 & -0.0312 + 0.0846i & 0.0312 + 0.0846i\\
7 & -1.3495 + 0.4494i & 1.3495 + 0.4494i\\
9 & -0.1133 - 0.3505i & 0.1133 - 0.3505i\\
11 & -0.8291 - 0.3872i & 0.8291 - 0.3872i\\
13 & -0.7264 - 0.4930i & 0.7264 - 0.4930i\\
\vdots & \vdots & \vdots\\
77 & -0.7494 - 0.5182i & 0.7494 - 0.5182i\\
79 & -0.7493 - 0.5182i & 0.7493 - 0.5182i\\
\hline
\text{exact (numerics)} & -0.7493 - 0.5182i & 0.7493 - 0.5182i
\end{tabular}
\caption{The longest-lived non-hydrodynamic mode pair extracted from the coefficients of the truncated hydrodynamic expansion of $\omega(q)$, compared to the exact value. The method is described in section \ref{extending}, for which we begin to obtain a reasonable estimate from hydrodynamic order $11$. \label{nonhydrotable}}
\end{center}
\end{table}

\flush
\section{Discussion\label{discussion}}
The phenomena of extracting excited states through analytic continuation from the ground state has a long history, to give two concrete examples, the quantum mechanical anharmonic oscillator \cite{PhysRev.184.1231} and obtaining first excited state from the Thermodynamic Bethe Ansatz in integrable two-dimensional field theories \cite{Dorey:1996re}\footnote{We thank an anonymous referee for pointing out this work.}. In this regard the analytic continuation we have performed demonstrates that the same approach can be applied to computing the QNMs of black holes and CFTs. A key result of this paper is that the analytic continuation was performed in hydrodynamics, that is, we needed only the Taylor expansion of the ground state QNM near the origin of the complex $q$ plane in order to access the excited states.\footnote{This procedure can also be performed directly in the holographic context by solving the QNM equations, for instance by using an existing QNM solution at momentum $q_{n}$ as a numerical seed for constructing another solution at $q_{n+1}$, one can iteratively follow a trajectory that encircles the branch points in the complex $q$ plane.}

By a suitable summation of the hydrodynamic Taylor expansion of the dispersion relation of the shear diffusion mode, we extracted the frequencies of non-hydrodynamic modes (i.e. the excited states). In particular at order 79 in the hydrodynamic expansion we extracted a pair of off-axis modes,
\be
\omega(0)/\mu = \pm 0.7493 - 0.5182i
\ee
in agreement with a pair of off-axis short-lived modes coming from the underlying holographic theory, here up to four significant digits. These are the longest lived non-hydrodynamic modes in this channel. Our method explored the analytic structure of $\omega(q)$ in complex $q$, switching complex variables to cover multiple sheets where branch points were encountered. 
We designed the map of complex variables to be as simple as possible, built from a quadratic relation in order to produce the two sheets at a square-root branch point. From there we found that summing the series as a Pad\'e approximant gave excellent results. 

We were able to infer where the branch points occurred empirically, by the convergence properties of the Taylor expansion itself. However, the results we obtained on the real axis did not appear to be particularly sensitive to the precise location of the $iq_\triangleleft$ branch point -- we demonstrated that the associated errors near this point, whilst very large, were damped as one approaches the real axis. This points to a simpler method to extract the modes, perhaps requiring only topological data.

\begin{figure}[h!]
\begin{center}
\includegraphics[width=0.75\columnwidth,clip]{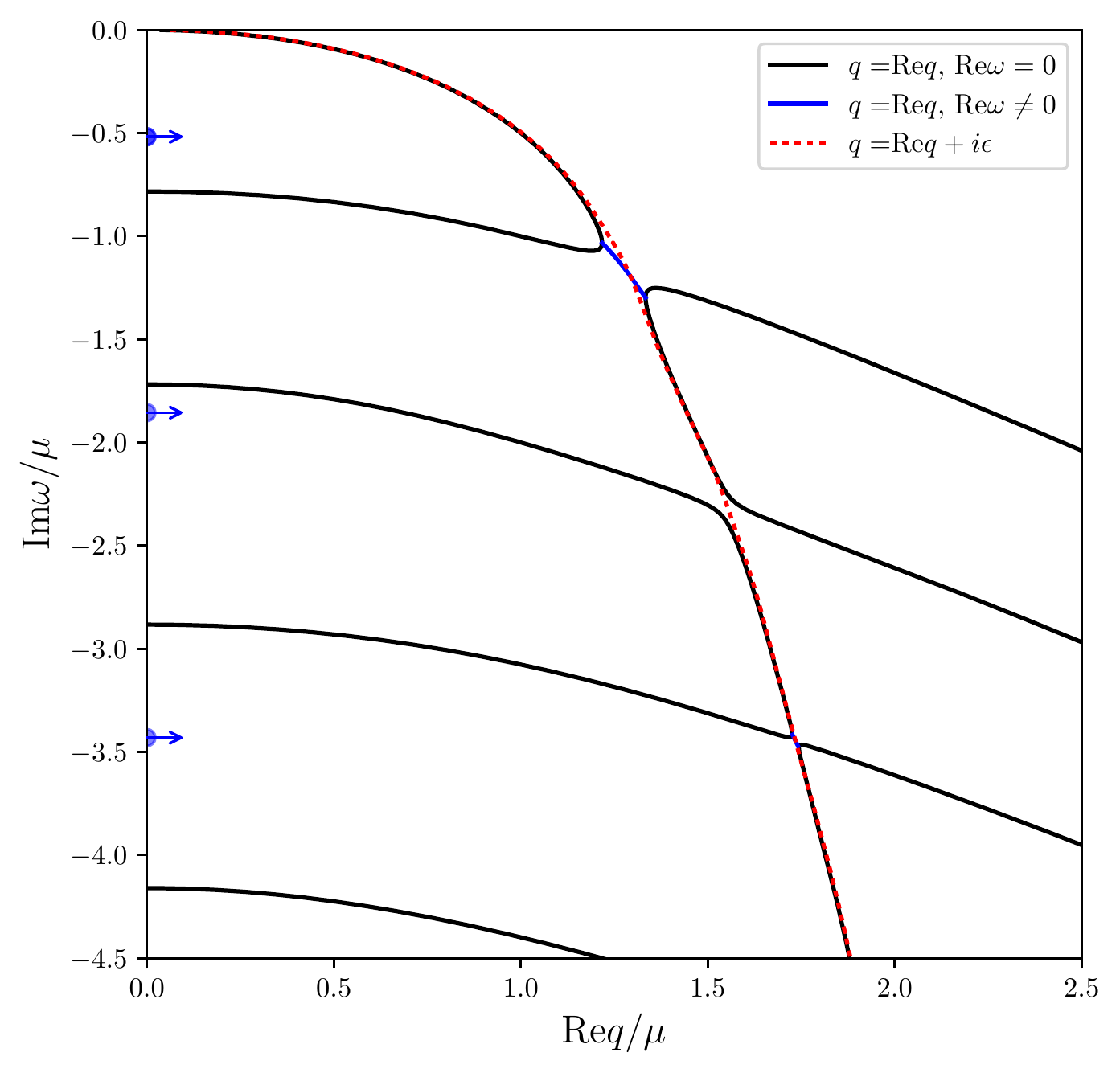}
\caption{The hydrodynamic highway at $T/\mu = 1/(4\pi)$: purely imaginary (black) and off-axis (blue) modes. The red dashed line shows a trajectory that stays on the hydrodynamic sheet, passing a series of branch cuts which may be used to exit to any of the short-lived quasinormal mode sheets. \label{highway}}
\end{center}
\end{figure}

We restricted our attention to the first few sheets, but it is an interesting question in general as to how much information about the spectrum of the underlying theory -- in this case the quasinormal mode spectrum of a black hole -- can be mined from the hydrodynamic data. It would of course also be desirable to have a systematic prescription to convert hydrodynamic data into a spectrum of non-hydrodynamic modes, perhaps one that does not rely on such a detailed understanding of the analytic properties, and the connectivity of the sheets in the complex $q$ plane. 
We postpone such questions to future work, except to say that at certain values of $T/\mu$ in this particular model, the level of branching can be very high, see figure \ref{highway}. At this value of $T/\mu$, along $q=$Re$q+i \epsilon$ on the hydrodynamic sheet, one can seemingly access infinitely many other sheets via a series of branch cuts further down the Im$\omega$ axis.

Relating the hydrodynamic expansion of dispersion relations considered here to non-perturbative corrections and the subsequent transseries structure of say, one-point functions, is another interesting direction. Some discussion along these lines can be found in \cite{Heller:2016gbp}, arguing that a dispersion relation with non-zero radius of convergence can lead to a factorial growth and a divergent series for a one-point function in real space. The analysis used here relied on branch points to get from one sheet to another, but for the model considered these are pushed to infinity as $\mu \to 0$. It would be interesting to understand if this has any bearing on whether or not certain transient transseries contributions appear.

Finally we conclude by observing that the structure of dispersion relations in the complex $q$ plane, which we exploited here, is not only of formal interest. Such modes can be physically relevant in situations where the resulting exponential growth in space is cut off by a physical source or boundary condition, as has been recently demonstrated in \cite{Sonner:2017jcf} for the case of steady flows over codimension-1 obstacles. In the relativistic context, instead of taking the Im$q=0$ slice of the dispersion relation as a surface in complex $\omega$ and complex $q$, these modes may be viewed as the result of a different slicing that is defined by the velocity of the background fluid.

\section*{Acknowledgements}
It is a pleasure to thank Massimiliano Ronzani, Julian Sonner, Peter Wittwer and Szabolcs Zakany for discussions.
I am supported by the NCCR under grant number 51NF40-141869 `The Mathematics of Physics' (SwissMAP).

\bibliographystyle{utphys}
\bibliography{shear}{}

\end{document}